\documentclass[a4paper,nofootinbib]{revtex4}
\usepackage[utf8]{inputenc}
\usepackage{lipsum}
\usepackage{xcolor}
\usepackage{amssymb,amsmath,amsthm,amsfonts,mathrsfs}
\usepackage{graphicx}
\usepackage{tabularx}
\usepackage{booktabs}
\usepackage{nameref}
\usepackage[hidelinks]{hyperref}
\usepackage[capitalise]{cleveref} 
\hypersetup{
    colorlinks=true,
    linkcolor=black,
    citecolor=blue,
    filecolor=black,
    urlcolor=blue,
    pdfauthor={M. Cadoni, L. Modesto, M. Pitzalis, and A. P. Sanna},
    pdftitle={Stable Wormholes in Conformal Gravity}
}

\usepackage{xcolor}
\usepackage{subfigure}
\usepackage[normalem]{ulem}
\newcommand{\be}{\begin{equation}}
\newcommand{\ee}{\end{equation}}
\newcommand{\bea}{\begin{eqnarray}}
\newcommand{\eea}{\end{eqnarray}}

\def\dd{\text{d}}

\def\lb{\label}

\newcommand{\bes}{\begin{split}}
\newcommand{\ees}{\end{split}}

\begin{document}
\title{\textbf{ Stable Wormholes in Conformal  Gravity}}
 \author{M. Cadoni$^{a,b}$}
\email{mariano.cadoni@ca.infn.it}

\author{L. Modesto$^{a,b}$}
\email{lmodesto@sustech.edu.cn}

\author{M. Pitzalis$^{a,b}$}
\email{mirko.pitzalis@ca.infn.it}

\author{A. P. Sanna${}^{a,b}$}
\email{asanna@dsf.unica.it}

\affiliation{$^{a}$Dipartimento di Fisica, Universit\`a di Cagliari, Cittadella Universitaria, 09042 Monserrato, Italy}
\affiliation{$^{b}$I.N.F.N, Sezione di Cagliari, Cittadella Universitaria, 09042 Monserrato, Italy}

\begin{abstract}
We present a class of Lorentzian traversable wormholes in conformal gravity, constructed via Weyl rescaling of Minkowski spacetime. As a result, these wormholes are solutions of every theory of gravity that is both conformally invariant and admits Minkowski spacetime as a solution. We specifically examine the case of a wormhole possessing a Morris-Thorne shape function, arising as a solution of a scalar-tensor conformally invariant theory of gravity. We show that these solutions represent regular, traversable wormholes that are also stable at the linear perturbation level. 
We argue that, when the Weyl symmetry is spontaneously broken, the broken symmetry phase may lead to a stable ``wormhole phase" alternative to the flat ``Minkowski phase".
\end{abstract}

\maketitle 

\tableofcontents

\section{Introduction}
Wormholes (WHs) are regular smooth solutions of Einstein-like theories of gravity connecting two different regions of spacetime, generally asymptotically-flat (AF) or (Anti) de Sitter [(A)dS], through an intermediate throat region~\cite{Morris:1988cz,Visser:1995cc}. In particular, traversable Lorentzian WHs allow timelike (null) signals emitted from one asymptotic region to reach the other one in finite proper (affine) time.

Initially proposed by Einstein and Rosen as an attempt to provide a geometrical description of elementary particles within general relativity (GR)~\cite{Einstein:1935tc}, WHs were later employed as an educational tool for teaching GR~\cite{Morris:1988cz}. Since then, Lorentzian traversable WHs have quickly evolved into intriguing objects of study in both fundamental theoretical physics and popular science. Their unique ability to connect two distinct regions of spacetime has made them relevant in various areas of gravitational physics. Specifically, WHs have been explored as a means to investigate fundamental aspects of spacetime, including causality~\cite{Ellis:1973yv}, rapid interstellar travel~\cite{Morris:1988cz}, the existence of closed timelike curves~\cite{Hawking:1991nk,Visser:1992tx}, superluminal propagation~\cite{Visser:1995cc}, and the physical interpretation of energy conditions~\cite{Bronnikov:1973fh,Visser:1989kh,Visser:1995cc,Hochberg:1998ha,Hochberg:1998ii,Lobo:2004rp,Lobo:2005us,Garattini:2007ff,Bouhmadi-Lopez:2014gza,Garattini:2019ivd,Lobo:2020ffi,Barros:2023pre,Garattini:2023kmr,Garattini:2024jkr}. More recently, they have gained attention for their potential role in addressing the black-hole information puzzle~\cite{Maldacena:2013xja,Penington:2019kki,Almheiri:2019qdq,Almheiri:2020cfm}. Furthermore, WHs have been proposed as black-hole mimickers (see, e.g., Refs.~\cite{Damour:2007ap,Cardoso:2016rao,Konoplya:2016hmd,Harko:2017fra,Bueno:2017hyj,Cardoso:2019rvt,Akil:2022coa} and references therein), offering alternative models for astrophysical observations. They also appear in some approaches to quantum gravity~\cite{Hawking:1988ae,Vollick:1998qf,Lobo:2007qi,Garattini:2007ff,Arkani-Hamed:2007cpn,Garattini:2008xz,Garattini:2011fs,Sengupta:2021wvi,Sengupta:2023yof}.

Despite their theoretical appeal, most known WH solutions suffer from two major drawbacks. First, they are typically derived using a reverse-engineering approach. In this method, one first postulates the spacetime geometry and then determines the stress-energy tensor of the (exotic) matter required to support the solution using Einstein’s equations. Only a few examples are known where the construction starts from a given form of the stress-energy tensor. Even in these cases, however, formulating an action for the coupled gravity-matter system remains highly challenging~\footnote{This is not entirely true for Euclidean WH solutions, where, in some specific cases, an action can indeed be formulated (see, e.g., Refs.~\cite{Cadoni:1994av,Cadoni:1995pg}.}. This limitation is particularly significant in the context of quantum gravity and other fundamental applications, where an explicit action would be highly desiderable. Second, GR WHs are typically unstable due to their exotic matter source~\cite{Gonzalez:2008wd,Bronnikov:2012ch}.

A framework that is largely unexplored but might be very promising for finding WH solutions is conformal gravity. This class of theories employ a well-defined action and is characterized by invariance under coordinate-dependent Weyl rescalings of the spacetime metric~\footnote{Examples of theories invariant under such a transformation will be provided in \cref{sect.implications}}
 

%
\be\lb{wrt}
g_{\mu \nu}\to g^{\prime}_{\mu \nu}=\Omega^2(x) g_{\mu \nu}\, ,
\ee
where $g_{\mu \nu}$ is a generic spacetime metric tensor and $g^{\prime}_{\mu \nu}$ is the Weyl rescaling of $g_{\mu \nu}$ with conformal factor $\Omega(x)$. A peculiarity of this framework is that the theory's broad symmetry allows for a wide range of possible solutions. Moreover, such a broad symmetry might also play a crucial role in stabilizing WH solutions.

Additionally, numerous indications suggest that gravity may exhibit a conformal phase in the deep ultraviolet regime~\cite{Reuter:2008qx,tHooft:2010xlr,tHooft:2010mvw,Mannheim:2011ds,Kallosh:2013lkr,tHooft:2015vaz,Rachwal:2018gwu,Modesto:2016max,Rachwal:2022pfe,Giacometti:2024qva}, which might play an important role in the ultraviolet completions of gravity. On the other hand, the conformal symmetry  must be (spontaneously) broken at low energies. The existence of WH solutions in the broken conformal phase would be particularly intriguing, especially in the context of quantum gravity and the black-hole information puzzle.
A first attempt to construct WH solution in the conformal gravity framework was made in~\cite{Hohmann:2018shl}, where it was shown that WH can exist without the need of exotic matter sources. However, this construction method employed conformal transformations of the Schwarzschild-(A)dS solution, which are  not suitable to generate AF WHs.\\

The goal of this paper is to explore the existence of AF WH solutions within conformal theories of gravity.
In \cref{sect.worm}, we explicitly construct an entire class of Lorentzian traversable WHs via a Weyl rescaling of Minkowski spacetime. A key implication of this result is that any gravity theory invariant under the Weyl rescaling \eqref{wrt} and admitting Minkowski spacetime as a solution must also inherently support this class of WHs. However, the Weyl-conformal symmetry raises conceptual issues regarding the physical interpretation of WH solutions within this theory. This is discussed in \cref{sect.implications}. Next, in \cref{sect.scalartensor}, we analyze in detail a well-known case: a conformally invariant scalar-tensor theory of gravity. Specifically, we focus on an example from our general class of WH solutions, where the shape function follows the Morris-Thorne form. We then examine key properties of these WHs, including their traversability (\cref{sect.traversability}), geodesic completeness of the spacetime (\cref{sect.geodesic}), the associated energy conditions (\cref{sect.energy}) and stability under linear perturbation (\cref{sect.stability}). We conclude the paper with some final considerations in \cref{sect.conclusions}.

\section{Generating wormhole solutions using a Weyl rescaling of Minkowski spacetime }
\label{sect.worm}

An AF traversable WH is a solution of a metric theory of gravity, which is characterized by a throat connecting two AF regions. The metric is usually assumed to be both spherically-symmetric and static~\footnote{This latter condition is not strictly necessary as the WH could be unstable,  but we will adopt it for simplicity. For stationary WH models, see, e.g., Refs.~\cite{Teo:1998dp,Kuhfittig:2003wr,Kashargin:2007mm,Kashargin:2008pk,Dzhunushaliev:2013jja,Kleihaus:2014dla,Franzin:2022iai,Garattini:2025gfq}.}. These symmetries imply that the WH spacetime metric must have the form~\cite{Morris:1988cz}~\footnote{In this paper, we adopt natural units, in which $c = \hbar = 1$.}
\be
\lb{whgeneral}
\dd s^2 = -e^{2\Psi(r)}\dd t^2+\frac{\dd r^2}{1-\frac{b(r)}{r}}+r^2 \dd \Omega_2^2\, ,
\ee
where $b(r)$ and $\Psi(r)$ depend on the radial coordinate $r$ only and are called  the shape and redshift functions, respectively.  We can also express the metric in terms of the proper radial distance from the WH throat $l$, defined as
\be\lb{prd}
l=\pm\int \frac{\dd r }{\sqrt{1-\frac{b(r)}{r}}},
\ee
where the $\pm$ sign refers to the upper and lower universe, respectively. In this coordinate system the metric reads
\be\label{properdistancemetricgeneral}
\dd s^{2} = - A(l^2) \dd t^{2} + \dd l^{2} + B(l^2) \dd\Omega_{2}^{2}\, ,
\ee
where $A(l^2)$ and $B(l^2)$ are metric functions.

Regularity and traversability of the WH impose constraints on the form of $b(r)$ and $\Psi(r)$, expressed by the following requirements:
\begin{itemize}
    \item[$(a)$]The redshift function $\Psi$ has to be always finite to prevent horizons or singularities. Additionally, the existence of two AF regions requires $\Psi\to 0$ as $l \to \pm \infty$;
    \item[$(b)$] The spatial geometry must exhibit a WH shape. Thus, equation $b(r) = r$ must have the solution $r = b_0$, which identifies the position of the throat. The latter, in turn, corresponds to the minimum of the proper distance $l(r)$;
    \item[$(c)$] The proper distance must be strictly increasing on both sides of the throat. This requirement is satisfied if the flaring out condition $b'(r=b_0)< 0$ holds;
    \item[$(d)$] A timelike/null signal originating from one of the AF regions should reach the throat at $r = b_0$ in finite proper (affine) time;
    \item[$(e)$] Curvature singularities must be absent and the spacetime must be geodesically complete.
\end{itemize}

To demonstrate how traversable WHs can be generated from flat Minkowski spacetime through the conformal transformation \eqref{wrt}, we begin by expressing Minkowski spacetime in spherical coordinates
\be\lb{min}
\dd s^2= - \dd t^2+\dd x^2+x^2 \dd \Omega_2^2  \, .
\ee
Then we perform a Weyl rescaling of the Minkowski metric, along with a reparametrization of the radial coordinate $x=x(r)$ as follows 
\be
\lb{WR}
g_{\mu \nu}\to g'_{\mu \nu}=Q^2(r) g_{\mu \nu}, \qquad   x=\frac{r}{Q(r)}\, .
\ee 
According to \cref{wrt}, if $g_{\mu\nu}$ is an exact solution of the conformal gravity theory, then $g^{\prime}_{\mu\nu}$ is also solution~\cite{Modesto:2021yyf,Bambi:2016wdn,Chakrabarty:2017ysw}. It is straightforward to verify that the transformation \eqref{wrt} brings the Minkowski metric into the WH form \eqref{whgeneral}, with the redshift and shape function determined completely in terms of $Q(r)$
\be
\lb{wc}
\left(1-r \frac{Q'}{Q}\right)^2= \left[1-\frac{b(r)}{r}\right]^{-1}\, , \qquad e^{2\Psi}= Q^2\, ,
\ee
where the prime denotes the derivation with respect to $r$.
Instead of fixing the form of $Q$ and computing the corresponding expressions of $b$ and $\Psi$, one can take the reverse approach. Specifically, by selecting a shape function $b(r)$ that defines a particular WH solution, one can use \cref{wc} to determine the rescaling function $Q$ and, ultimately, the redshift $\Psi$.
Since the WH profile is given, conditions $(b)$ and $(c)$ are automatically satisfied. Meanwhile, $Q$ (and consequently $\Psi$) is determined by solving the first-order ODE in \cref{wc}. If the integration constant can be determined to satisfy conditions $(a)$, $(d)$ and $(e)$, a well-defined traversable WH solution is generated.
This procedure defines a broad class of traversable WH solutions that can be derived through a conformal rescaling of the Minkowski spacetime. In the following sections, we will present an explicit example of this procedure and thoroughly examine its features. 

\section{Wormhole solutions in conformal theories of gravity}
\label{sect.implications}

An important consequence of the results of the previous section is that any conformal theory of gravity that admits flat Minkowski spacetime as a solution must also inherently allow for the class of WH solutions presented above.

The simplest and most commonly used conformal theories of gravity are Weyl gravity and conformally invariant scalar-tensor gravity. The former is described by the action
\be\lb{wa}
S_\text{W}= \alpha \int \dd^4 x \, \sqrt{- g}\,  C_{\alpha \beta \gamma \nu} C^{\alpha \beta \gamma \nu} \, ,
\ee
where $C_{\alpha \beta \gamma \nu}$ is the Weyl tensor and $\alpha$ is a dimensionless  constant. This action is equivalent, up to boundary terms, to the quadratic action~\cite{Bach:1921zdq}
\be\lb{qua}
S_\text{Q}= \alpha_\text{q} \int \dd^4 x \, \sqrt{- g}\, \left( R_{\alpha \beta} R^{\alpha \beta } -\frac{1}{3} R^2\right)\, .
\ee
Both actions are invariant under the Weyl rescaling of the metric \eqref{wrt} and admit the Minkowski solution. Consequently, they must also include the class of WH solutions constructed in \cref{sect.worm} among their possible solutions.
However, the conformal gravity theories \eqref{wa} and \eqref{qua} exhibit an undesirable feature in the context of WHs. Although they allow for the Minkowski and Schwarzschild solutions \cite{Mannheim:1988dj}, there is no simple way to reduce their field equations to the usual Einstein ones of GR (see however Ref.~\cite{Wheeler:2013ora}). As a result, it is difficult to identify a well-defined stress-energy tensor that serves as the source for the WH solution, making the discussion of energy conditions particularly problematic.

In contrast, the conformal scalar-tensor theory of gravity is free from this difficulty. The action is formulated in terms of the metric and a non-minimally coupled scalar field  $\Phi$
\label{sect.scalartensor}
\be
\lb{actiongeneral}
S= \int \dd^4 x \, \sqrt{g} \, \left( \Phi^2 R+ 6 g^{\mu \nu} \partial_\mu \Phi  \partial_\nu \Phi \right)\, .
\ee
Originally, in Weyl's theory, $\Phi$ was introduced as a field providing an arbitrary  local unit of mass.  However, we can also treat $\Phi$  as a coupling, more precisely we can express it in terms of a coordinate-dependent Newton constant
\be
\lb{rnc}
\Phi^2= \frac{1}{16\pi G}\, ,
\ee
In the following, we will adopt this interpretation.

The action is invariant under the Weyl rescaling of the metric and the scalar field~\footnote{One can also add to the action a self-interaction, Weyl symmetry preserving potential term $\lambda \Phi^4$. However, for $\lambda \neq 0$, the Minkowski solution is not allowed.}
\be
\lb{ct}
g_{\mu \nu}\to g'_{\mu \nu}=\Omega^2 g_{\mu \nu}, \quad \Phi\to \Phi'= \Omega^{-1}\Phi.
\ee
The scalar field is a pure gauge degree of freedom. Indeed, choosing $\Omega = \lambda \Phi$ (with $\lambda$ a dimensionful constant) any solution of the theory can be mapped to an equivalent solution with a constant scalar field $\Phi = 1/\lambda$. 
The field equations stemming from the action \eqref{actiongeneral} are
\begin{subequations}
\begin{align}
    \Phi^2 G_{\mu \nu} &= \nabla_\mu\nabla_\nu \Phi^2 - g_{\mu \nu} \Box \Phi^2 -6 \left( \partial_\mu \Phi  \partial_\nu \Phi -\frac{1}{2} g_{\mu \nu}  (\partial \Phi)^2\right) \, ; \label{fe}\\
    \Box \Phi &= \frac{1}{6}\Phi R\, .
\label{KGgeneral}
\end{align}
\end{subequations}
These equations allow for the Minkowski vacuum solution endowed with a constant scalar field $\Phi=\Phi_0=1/\sqrt{16\pi G_\text{N}}$~\footnote{For clarity, we use different notations for the coordinate dependent Newton coupling $G$, as in \cref{rnc}, and the usual Newton constant $G_\text{N}$.}. 
Following the method outlined in \cref{sect.worm}, we can construct the WH solution \eqref{whgeneral} through \cref{wc}. A crucial difference is that, now, the conformal transformation \eqref{ct} produces a non-trivial scalar field, i.e., a WH dressed with the scalar-field configuration 
\be\lb{sfc}
\Phi = Q^{-1} \Phi_0\, .
\ee
Furthermore, the field equations can be written in the Einstein form $G_{\mu\nu}= 8\pi G_\text{N} T_{\mu\nu}$, with the effective stress-energy tensor of the scalar field given by 
\be\lb{set}
T_{\mu\nu}=\frac{1}{8 \pi G_\text{N}\Phi^2 } \left[\nabla_\mu\nabla_\nu \Phi^2 - g_{\mu \nu} \Box \Phi^2 -6 \left( \partial_\mu \Phi  \partial_\nu \Phi -\frac{1}{2} g_{\mu \nu}  (\partial \Phi)^2\right)\right].
\ee
This can be effectively considered as the source of the WH solution. Notably, \cref{set} is the covariantization of the new improved stress-energy tensor for a conformal scalar field in flat space, proposed in Ref.~\cite{CALLAN197042} to ensure finiteness at all orders in renormalized perturbation theory. On shell its trace is 
\begin{equation}
    T^\mu_\mu = -\frac{6}{8\pi G_\text{N} \Phi} \Box \Phi\, ,
    \label{traceSET}
\end{equation}
making it traceless whenever the spacetime satisfies $R=0$ (see \cref{KGgeneral}).   

Physically, identifying \cref{set} as the WH source connects directly to interpreting $\Phi$ as a coordinate-dependent Newton constant, a point we now discuss. 

\subsection{Physical meaning of conformal frames}
\label{sect.physical}

The Weyl rescaling \eqref{WR}, which relates our WH solution to Minkowski spacetime, represents a local gauge transformation, implying that these two spacetimes provide different yet equivalent descriptions of the same solution. Physically, these descriptions correspond to two observers adopting different ``gauges" to measure distances and times~\footnote{Historically, Weyl was the first to use the term ``gauge" in reference to local symmetries in the context of conformal gravity.}. 
In the scalar-tensor theory \eqref{actiongeneral}, a solution consists of both the metric tensor $g_{\mu\nu}$ and a scalar field $\Phi$, the latter acting as a pure gauge degree of freedom. Since the scalar field is non-minimally coupled, the theory falls within the Brans-Dicke framework, allowing $\Phi$ to be interpreted as a coordinate-dependent Newton constant (see \cref{rnc}). This leads to an alternative interpretation of our WH solution: Minkowski spacetime with a constant Newton coupling $\Phi_0$ is equivalent, in another conformal frame, to a curved spacetime--the WH solution--with a coordinate-dependent gravitational coupling.

These considerations hold as long as conformal symmetry remains unbroken. However, conformal Weyl symmetry may only be a fundamental symmetry of gravity at energy scales near the Planck mass and must be broken somehow at lower energies. The precise mechanism of this symmetry breaking remains unclear, though a compelling scenario is spontaneous symmetry breaking, which would determine both the background  metric $g^{(0)}_{\mu\nu}$ and the value of $\Phi_{0}$. The simplest symmetry-breaking background is the Minkowski vacuum \eqref{min} with a constant $\Phi_0$, which sets the Newton constant $G_\text{N}$. However, the WH solution \eqref{wh1} may represent an alternative phase of spontaneous conformal symmetry breaking, where a curved background geometry emerges alongside a coordinate-dependent Newton coupling. In this ``WH phase", the scalar field itself acts as the source of the WH. 

\section{A Morris-Thorne-like wormhole solution of conformal scalar-tensor gravity}
\label{sect.scalartensor}

To be more concrete, in the remainder of the paper, we will focus on a simple yet particularly insightful case within the broader class of WH models. Specifically, we consider solutions in which the spatial sections of the WH geometry correspond to the well-known Morris-Thorne model~\cite{Morris:1988cz}. The shape function is 
\be\lb{sf}
b(r)= \frac{b_0^2}{r}\, .
\ee
Following the procedure outlined at the end of \cref{sect.worm}, determining the redshift function requires first solving the ODE that relates $b(r)$ to the conformal factor $Q(r)$ (left equation in \cref{wc}). Choosing the negative root when solving for $Q'$ and fixing the integration constant such that $Q\to 1$ for $r\to \pm \infty$ (ensuring asymptotic flatness), one easily finds
\be\lb{eq7}
Q(r) = \frac{2}{1+\sqrt{1-\frac{b_0^2}{r^2}}}\, .
\ee
The traversable WH solution, then, reads as 
\be\lb{wh}
\dd s^2 = - 4 \left(1+\sqrt{1-\frac{b_0^2}{r^2}}\right)^{-2} \dd t^2+\frac{1} {1-\frac{b_0^2}{r^2}}\dd r^2+r^2 \dd \Omega_2^2\, , \qquad  \Phi=\frac{\Phi_0}{2}\left(1+\sqrt{1-\frac{b_0^2}{r^2}}\right)\, .
\ee
We can also write the metric using the proper radial distance from the throat  $l$. From  \cref{prd}, we get
\be
l=\int \frac{\dd r \, r}{\sqrt{r^2-b_0^2}} = \sqrt{r^2-b_0^2}\, .
\ee
\Cref{wh} is, then, recast as in \cref{properdistancemetricgeneral}, with
\be\lb{wh1}
A(l^2) = - 4 \left(1+\frac{b_0^2}{l^2} \right) \left(1+\sqrt{1+\frac{b_0^2}{l^2} } \right)^{-2} \, , \qquad B(l^2) = l^2 + b_0^2\, ,
\ee
together with the scalar field reading as
\begin{equation}
    \Phi= \frac{\Phi_0}{2}\left(1+\frac{1}{\sqrt{1+\frac{b_0^2}{l^2}}}\right)\,  .
\end{equation}
To confirm that this truly represents a traversable wormhole, we must verify the validity of conditions $(a)$ to $(e)$ outlined in \cref{sect.worm}.

Regarding condition $(a)$, the redshift function $\Psi$ in \cref{wh} is finite and nonzero everywhere. It is monotonically decreasing with $r$, ranging from $\ln 2$ at the throat ($r = b_0$ or, equivalently, $l \to 0$), to $0$ in the two AF regions at $r \to \pm \infty$ ($l \to \pm \infty$).  

On the one hand, conditions $(b)$ and $(c)$ are satisfied by construction, as the shape function is chosen to be equivalent to the Morris-Thorne model (see Ref.~\cite{Morris:1988cz}). 

On the other hand, conditions $(d)$ and $(e)$ require explicit verification. This is not only due to the different redshift function compared to the Morris-Thorne WH, but also to the fact that we are working within the framework of conformally invariant scalar-tensor gravity. In this setting, timelike particles can couple to the scalar field $\Phi$. Thus, conditions $(d)$ and $(e)$ will be examined in detail in the next two subsections.

\subsection{Wormhole traversability}
\label{sect.traversability}

The WH traversability can be assessed by investigating the geodesic motion of both timelike and null particles. 

Conformal transformations preserve the light-cone structure of spacetime, ensuring that null geodesics in our WH spacetime follow the same trajectories as in Minkowski spacetime. As a result, null particles traveling along geodesics can traverse the WH unobstructed.

This is not the case for timelike observers. Their motion can be characterized by the radial velocity measured by static observers, $v_l= \dd l/\dd s$, and the proper time $\Delta\tau$ required to reach the throat at $r = b_0$ from an initial position at radial coordinate $r_0$ in one of the two spacetime regions. Moreover, in scalar-tensor conformal gravity, an additional complication arises compared to standard GR WHs. A point-like particle does not only couple to the metric tensor in the usual minimal covariant way, but also interacts with the scalar field $\Phi$. Such coupling can be generally described in terms of a coupling function $F(\Phi)$. 
The particle action then reads
\be\lb{pcoupling}
S= -\int \dd \lambda \, \sqrt{-F^2(\Phi) \, g_{\mu \nu}\dot x^\mu\dot x^\nu}, 
\ee
where the dot denotes derivation with respect the affine parameter $\lambda$. To prevent singularities in the coupling, we assume that $F^2(\Phi)$ remains finite and nonzero for all finite values of $\Phi$. For generic $F^2(\Phi)$, the action \eqref{pcoupling} is not conformally invariant. The coupling becomes conformal, i.e., invariant under the Weyl rescaling \eqref{ct}, when $F(\Phi)=f \Phi$, with $f$ a constant.

Next, we compute the radial velocity $v_l$. Since our solutions are static, time-translation invariance ensures the existence of a conserved quantity $E$. Applying the Euler-Lagrange equations together with the timelike gauge condition $\dot x_\mu \dot x^\mu= -1$, we get
\be\lb{ELE}
\dot t^2= \frac{E^2}{F^2(\Phi) g_{00}^2}\, , \qquad \dot r^2= g^{rr} \left(\frac{E^2}{F^2(\Phi) |g_{00}|}-1\right)=g^{rr}\left(\epsilon^2(\Phi)-1\right)\, ,
\ee
where in the last step we used $g_{00}=-\Phi_0^2/\Phi^2$ and we have defined 
\be\lb{EP}
\epsilon(\Phi) \equiv \frac{E \Phi }{ F(\Phi)\Phi_0}\, .
\ee
From \cref{ELE}, the motion of the particles is physical as long as $\epsilon(\Phi)^2\ge 1$.

Using \cref{ELE} we can easily obtain $v^2_l$
\be\lb{ELE1}
v^2_l= \frac{\dot r^2}{\dot t^2}\frac{g_{rr}}{ |g_{00}|}= 1-\frac{F^2(\Phi)\Phi_0^2}{\Phi^2E^2}=1- \epsilon^{-2}(\Phi). \,
\ee
The observer's speed during radial motion remains finite and always less than the speed of light. For generic coupling, $v_l$ depends on the scalar field $\Phi$, but it is solely a function of $l^2$ (see \cref{wh1}). As a result, geodesics can therefore be continued at negative values of $l$, making the WH fully traversable.

As expected, when the coupling is conformal, i.e., $F(\Phi)=f \Phi$, the parameter $\epsilon$ becomes constant $\epsilon=\epsilon_0 \equiv E/f \Phi_0$, so that $v_l$ no longer depend on the scalar field but only on its asymptotic value $\Phi_0$. By setting $f=\Phi_0=1$, the observer's speed matches that of Minkowski spacetime, consistently with the fact that our WH solution is conformally related to the Minkowski spacetime. In this case, we can easily integrate \cref{ELE1} to get the geodesic equation 
\be
l(\lambda)= \sqrt{1-\epsilon_0^{-2}} \, \,\lambda + \text{constant} \, ,
\ee
which indeed describes a straight line.

The proper time along a radial geodesic needed to reach the throat at $r=b_0$ from a generic external point $r_0$ is, inverting \cref{ELE},
\be\lb{ELE2}
\Delta \tau(r_0)=\int_{r_0}^{b_0}  \frac{\dd r}
{\sqrt{\left(1-\frac{b_0^2}{r^2}\right)\left(\epsilon^2(\Phi)-1\right)}}\, .
\ee
For a general coupling function $F(\Phi)$, $\Delta \tau$ cannot be computed analytically. However, the regularity of $F(\Phi)$ ensures that $\epsilon^2(\Phi)$ remains well-behaved. Consequently, the only potentially divergent contribution to the integral \eqref{ELE2} may arise from the denominator evaluated near $r=b_0$. This contribution behaves as $\sim \sqrt {r^2-b_0^2}$, which vanishes at $r = b_0$. Therefore, $\Delta \tau$ remains finite, confirming that the WH is traversable within a finite proper time.

The integral in \cref{ELE2} can be explicitly evaluated when the coupling is conformal $F= f \Phi$ ($\epsilon= \epsilon_0 = \text{constant}$), to yield
\be\lb{ELE3}
\Delta \tau(r_0)=\sqrt{\frac{r_0^2-b_0^2}{\epsilon_0^2-1}}\, ,
\ee
which again remains finite, confirming the WH traversability.

\subsection{Regularity of the spacetime and geodesic completeness}
\label{sect.geodesic}

\begin{figure}[!ht]
    \centering
    \includegraphics[width=0.65\linewidth]{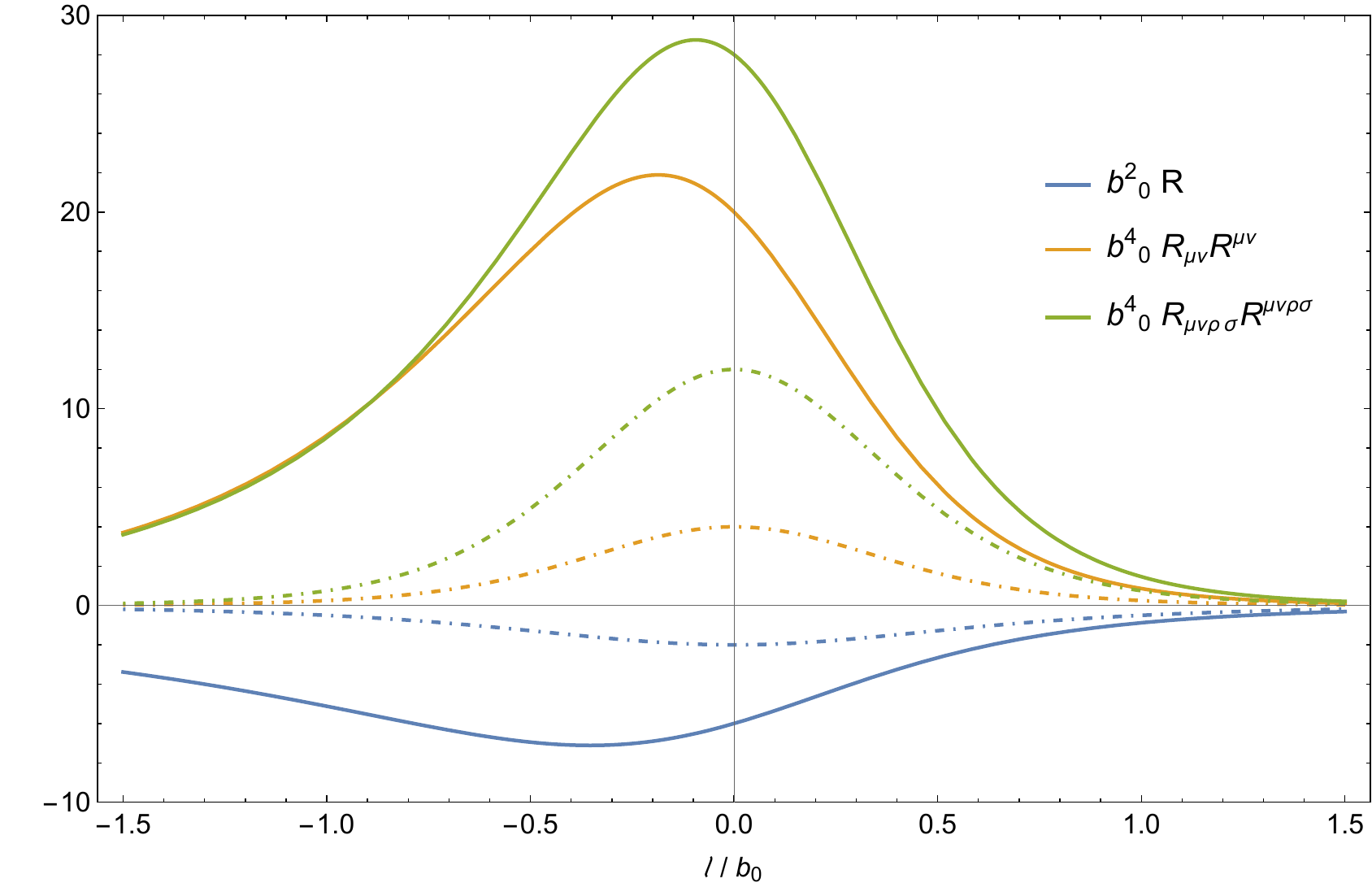}
    \caption{Curvature invariants for the wormhole metric \eqref{wh1} as a function of $l / b_0$ (solid lines), compared with the same quantities evaluated with the Morris-Thorne metric (dotdashed lines) which are characterized by  a zero redshift function.}
    \label{fig:curvatureinvariants}
\end{figure}

In GR, one way to probe the regularity of spacetime is by examining the behavior of  curvature invariants (i.e., quantities invariant under the $\text{Diff}$ group). However, for theories invariant under Weyl rescaling, the situation is more involved. The usual curvature invariants lack gauge-invariant meaning since they are not invariant under the full $\text{Diff} \times \text{Weyl}$ group. As long as Weyl symmetry remains unbroken, the regularity of spacetime should be characterized using quantities that are invariant under both diffeomorphisms and Weyl transformations.
Conversely, since the Weyl symmetry must be broken to produce the observable universe, the usual curvature invariants can be used to asses the regularity of the spacetime. 
Given the conformal relation with Minkowski spacetime, the analysis of the regularity of the WH geometry  is straightforward. As for the $\text{Diff} \times \text{Weyl}$ invariants, they must necessarily involve a combination of curvature tensors and derivatives of the scalar field. A notable example of such invariants is the Lagrangian in \cref{actiongeneral}. 
These invariants can be easily computed in the ``Minkowskian" conformal frame, where they identically vanish. In broken conformal phase, one can consider the usual curvature invariants: $R$, $R_{\mu\nu} R^{\mu\nu}$, $R_{\mu\nu \sigma \rho} R^{\mu \nu \sigma \rho}$ and $C_{\mu\nu\rho\sigma}C^{\mu\nu\rho\sigma}$. They are trivially zero in the ``Minkowski phase", whereas 
for the WH geometry \eqref{wh1}, they remain finite everywhere. In  \cref{fig:curvatureinvariants}, we show the behavior of the first three curvature invariants~\footnote{The contraction of the Weyl tensor can be written as a linear combination of the other invariants and, thus, it is not plotted in \cref{fig:curvatureinvariants}.} as a function of the proper distance $l$, evaluated for the solution \eqref{wh1}. They are also compared to the corresponding quantities for the Morris-Thorne WH. Notice that, unlike the latter model, in our case, the curvature invariants are not symmetric under the $l\to-l$ transformation. This is due to the presence of a non-constant redshift function. \\

Another way to assess the regularity of spacetime is by examining null and timelike geodesics. Pathological points correspond to locations where geodesics either terminate at a finite proper length or focus into caustic. As discussed in \cref{sect.traversability}, timelike radial geodesics originating from one asymptotic region reach the throat $b_0$ in finite proper time, cross it, and emerge in the other asymptotic region, confirming the geodesic completeness of the wormhole geometry \eqref{wh1}.
In principle, this argument does not rule out the possibility of geodesic focusing and caustic formation. While geodesic focusing may occur in certain spacetime regions, the formation of caustics is prevented by a de-focusing mechanism. As we will show in \cref{sect.energy}, this de-focusing arises due to the violation of the null energy condition. Indeed, explicit computations of the geodesic expansion parameter $\Theta$ in the WH geometry \eqref{wh1} confirm that caustics never form.

\subsection{Energy conditions}
\label{sect.energy}

Let us now investigate the energy conditions of our WH solution. In general, we expect a violation of the null energy condition, as this is typically required for WH traversability.
Energy conditions can be investigated by computing the components of the effective stress-energy tensor \eqref{set}, which can be interpreted as a relativistic anisotropic fluid described by 
\begin{equation}
    T_{\mu\nu} = \left(\rho + p_\perp \right) u_\mu u_\nu + p_\perp g_{\mu\nu} -\left(p_\perp - p_\parallel \right) w_\mu w_\nu \, .
    \label{anisoSET}
\end{equation}
$\rho$ is the energy density, $p_\parallel$ and $p_\perp$ the two pressure components, while $u_\mu$ and $w_\mu$ are, respectively, a time-like and space-like 4-vectors satisfying $u_\mu u^\mu = -w_\mu w^\mu = -1$ and $u_\mu w^\mu = 0$. Specifically, we have
\begin{equation}
    T^0_0 = -\rho \, , \qquad T^r_r = p_\parallel \, , \qquad T^\theta_\theta = p_\perp\, .
    \label{componentsanisoset}
\end{equation}
In terms of these components, the energy conditions read as follows:
\begin{itemize}
    \item Weak Energy Condition (WEC): \, $\rho \geq 0 \, \, \land \, \, \rho + p_\parallel \geq 0 \,\, \land \,\, \rho + p_\perp \geq 0$;
    \item Strong Energy Condition (SEC): \, $\rho + p_\parallel \geq 0 \, \, \land \, \,  \rho + p_\perp \geq 0 \, \, \land \,\, \rho + p_\parallel + 2p_\perp \geq 0$;
    \item Null Energy Condition (NEC): \,\, \, $\rho + p_\parallel \geq 0 \, \, \land \, \, \rho + p_\perp \geq 0$.
\end{itemize}
A NEC violation is sufficient to violate the other energy conditions, as it is contained in both the WEC and SEC. \\

Using \cref{set} and the field equation \eqref{KGgeneral}, we now compute the effective energy density and pressure components sourcing the WH solution. We have for the energy density 
\begin{equation}
    -\rho =T_0^0 = \frac{1}{8\pi G_\text{N} \Phi^2} \left[g^{rr} \Phi'^2 - 2\Phi \Box \Phi - 2 g^{00} \Gamma^r_{00} \Phi \Phi' \right]\, ,
\end{equation}
which simplifies to, upon using \cref{wh}
\begin{equation}
    8\pi G_\text{N}\rho = -\frac{b_0^2}{r^4}\, .
    \label{effectiverho}
\end{equation}
As expected from WH traversability, the energy density is always negative  so that the  WEC is violated everywhere in the WH spacetime. We also notice that the $r^{-4}$ scaling of the density profile is typical of conformal field theories in four dimensions, a direct consequence of the conformal symmetry. We will comment further on this below. 

\begin{figure}
    \centering
    \includegraphics[width=0.65\linewidth]{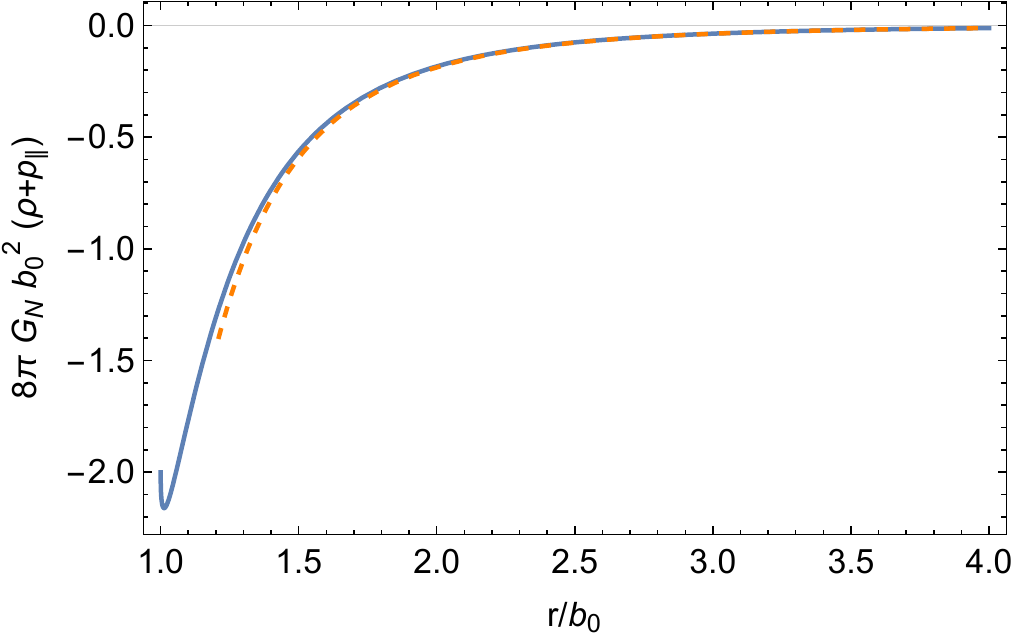}
    \caption{Behavior of $8\pi G_\text{N} b^2_0 \left(\rho + p_\parallel \right)$ as a function of $y \equiv r/b_0$ (solid blue line), showing the violation of the NEC. The orange dashed line represents the $-3/y^4$ asymptotic behavior. The $x$-axis has been cut at the throat $r = b_0$.}
    \label{fig:NEC}
\end{figure}

The radial pressure component, instead, is given by
\begin{equation}
 p_\parallel = T^r_r = \frac{1}{8\pi G_\text{N} \Phi^2} \left[2 g^{rr} \Phi \, \Phi'' - 3 g^{rr} \Phi'^2 - 2 g^{rr} \Gamma^r_{rr} \, \Phi \, \Phi' - 2\Phi \Box \Phi \right]\, ,
\end{equation}
leading to (for simplicity, we rescale the radial coordinate by defining $y \equiv r/b_0$)
\begin{equation}
    8\pi G_\text{N} p_\parallel = \frac{-4 \left(\sqrt{1-\frac{1}{y^2}}+1\right) y^2+3 \sqrt{1-\frac{1}{y^2}}+4}{b_0^2 \left(\sqrt{1-\frac{1}{y^2}}+1\right)^2 \sqrt{1-\frac{1}{y^2}} \, y^6}\, .
    \label{pparalleleffective}
\end{equation}
At the throat $r = b_0$ ($y = 1$), $8\pi G_\text{N} p_\parallel = -1/b_0^2$, so, as expected, the NEC (and, thus, the other energy conditions) is violated at this point. Actually, $\rho + p_\parallel < 0$ everywhere at finite $r$ (solid blue line in \cref{fig:NEC}). At $r \to \infty$ ($y \to \infty$), $8\pi G_\text{N} p_\parallel \sim -2 b_0^2/r^4$, so $8\pi G_\text{N}(\rho + p_\parallel) \to -3/b_0^2 r^4$ (orange dashed line in \cref{fig:NEC}), and it goes to zero at infinity, where $\rho + p_\parallel = 0$. This is sufficient to conclude that the NEC is violated everywhere, except at $r \to \infty$.

Finally, the tangential pressure reads
\begin{equation}
    p_\perp  = T^\theta_\theta = \frac{1}{8\pi G_\text{N} \Phi^2}\left[-2g^{\theta\theta} \Gamma^r_{\theta\theta} \Phi \, \Phi' + g^{rr} \Phi'^2 - 2\Phi \Box \Phi\right]\, ,
\end{equation}
leading to (using, as before, $y \equiv r/b_0$)
\begin{equation}
    8\pi G_\text{N} p_\perp = \frac{\left(4 \sqrt{1-\frac{1}{y^2}}+4\right) y^2-\sqrt{1-\frac{1}{y^2}}-4}{b_0^2 \left(\sqrt{1-\frac{1}{y^2}}+1\right)^2 \sqrt{1-\frac{1}{y^2}} y^6}\, .
    \label{pperpeffective}
\end{equation}
At the throat, $8\pi G_\text{N} p_\perp = 3/b_0^2>0$, while at spatial infinity, $8\pi G_\text{N} p_\perp = 2b_0^2/r^4$, again positive.

Thus, the energy conditions are satisfied only asymptotically, as it should be expected from AF traversable WHs. 

All the stress-energy tensor components feature a $1/r^4$ fall at infinity, typical of conformal fields in four dimensions. However, the stress-energy tensor \eqref{set} is not compatible with a conformal field theory, as its trace is not identically zero, but given by \cref{traceSET}~\footnote{One can also verify that \cref{traceSET} is consistent with $T^\mu_\mu = -\rho + p_\parallel + 2 p_\perp$ computed from \cref{anisoSET} and evaluated using \cref{effectiverho,pparalleleffective,pperpeffective}.}. As discussed in \cref{sect.implications}, using the field equation \eqref{KGgeneral}, $T^\mu_\mu = 0$ only for $R= 0$. In the WH spacetime, this condition is fulfilled only at $r \to \pm \infty$, where the action \eqref{actiongeneral} describes a conformal scalar field theory in Minkowski spacetime.
As expected, in the WH interior, the non-zero curvature of the spacetime breaks the conformal invariance of the asymptotic theory.

\subsection{Stability}
\lb{sect.stability}

In the GR framework, traversable WH are typically unstable. This instability arises from the necessity of exotic matter sources that violate the NEC in certain regions of spacetime to prevent the collapse of the WH throat. However, in our conformal gravity framework, this fate can be avoided due to the conformal symmetry of the theory. Although spontaneously broken, this symmetry can provide  a protection  mechanism that stabilizes the solution against instabilities at the linear perturbation level.

Consider perturbations of the WH background \eqref{wh}
\begin{equation}
    g_{\mu \nu}= g_{\mu \nu}^{\text{(WH)}} + h_{\mu\nu}^{\text{(WH)}}\, , \qquad \Phi= \Phi^{\text{(WH)}} + \phi^{\text{(WH)}} \, ,
    \label{pert}
\end{equation}
where $(g_{\mu \nu}^{\text{(WH)}}, \Phi^{\text{(WH)}} )$ is the background WH solution \eqref{wh} and $|h_{\mu\nu}^{\text{(WH)}}| \ll |g_{\mu \nu}^{\text{(WH)}}|$, $\phi^{\text{(WH)}} \ll \Phi^{\text{(WH)}}$ are the perturbations. 

Similarly, we consider perturbations of the conformally related Minkowski solutions
\begin{equation}
    g_{\mu \nu}= g_{\mu \nu}^{\text{(M)}} + h_{\mu\nu}^{\text{(M)}},\quad \Phi= \Phi_0 + \phi^{\text{(M)}} .
    \label{pert1}
\end{equation}
It has been shown by Percacci \cite{Percacci:2011uf} that the Weyl conformal symmetry \eqref{ct} of the scalar-tensor gravity theory \eqref{actiongeneral} remains preserved even at the level of linear perturbations. This implies that when perturbing the Minkowski solution, we can always ``gauge away" the scalar field perturbation in \cref{pert1}, setting $\phi^{\text{(M)}}=0$. Since the perturbations $\phi^{\text{(WH)}}$ and $\phi^{\text{(M)}}$ are related by the conformal Weyl rescaling \eqref{ct}, it follows that if $\phi^{\text{(M)}}=0$, then $\phi^{\text{(WH)}}$ must also vanish.
Furthermore, for the same reason, a small perturbation of the Minkowski solution \eqref{pert1} is mapped by the conformal transformation into a small perturbation of the WH background \eqref{pert}. Since we can ``gauge away" the perturbation of the scalar $\Phi^{\text{(WH)}}$, the equation governing the metric perturbation $h_{\mu\nu}^{\text{(WH)}}$ in \cref{pert} at the linear perturbation level coincides with that of Minkowski spacetime up to a rescaling. Consequently, the stability of the WH solution follows directly from the linear stability of Minkowski spacetime.

\section{Closing remarks}
\label{sect.conclusions}

In this paper, we have demonstrated that gravity theories invariant under a Weyl rescaling of the metric and admitting Minkowski spacetime as a solution always allow for a broad class of traversable and stable WH solutions. This result was established in a simple and general manner by constructing the WH through a Weyl rescaling of the Minkowski metric.

In particular, we analyzed in detail a specific case: a Morris-Thorne-like solution within the Einstein conformally invariant scalar-tensor theory. This framework is particularly intriguing because the scalar field can be interpreted both as a coordinate-dependent Newton coupling or as a component of the stress-energy tensor sourcing the WH. As expected, this stress-energy tensor violates the Null Energy Condition. However, this violation is less problematic than in conventional WH solutions, since, in our scenario, the scalar field is not an exotic matter source, but rather a pure gauge degree of freedom. Furthermore, the Weyl rescaling symmetry ensures that the WH solution inherits linear stability from the flat Minkowski background.

The existence of stable, traversable AF WH solutions in conformal gravity is a significant result. It suggests that the major challenges faced by WH solutions in General Relativity—such as the need for exotic matter and inherent instability—can be overcome by drastically enhancing the symmetries of the gravitational theory.

Our framework becomes even more compelling in the presence of spontaneous breaking of Weyl conformal symmetry. This could give rise to two distinct {\sl{stable}} phases of broken conformal symmetry: one corresponding to Minkowski spacetime with the standard Newton constant $G_\text{N}$, and another -- a ``WH phase" -- in which the geometry is described by our WH solution with a coordinate-dependent Newton coupling. 

There are strong indications that conformal symmetry could play an important role in the ultraviolet behavior of gravity. However, for consistency with observations, this symmetry must be broken—most likely spontaneously—at low energies to recover Einstein’s equations of motion. The possibility of a WH phase emerging from conformal symmetry breaking is an exciting possibility with potential implications for various fundamental issues in gravitational physics. These include the black-hole information problem (in particular in the context of the ER = EPR conjecture~\cite{Maldacena:2013xja}), AdS quantum gravity~\cite{Penington:2019kki,Almheiri:2019qdq,Almheiri:2020cfm,Marolf:2021kjc} and other approaches to quantum gravity~\cite{Hawking:1988ae,Vollick:1998qf,Arkani-Hamed:2007cpn,Garattini:2008xz,Garattini:2011fs,Sengupta:2021wvi,Sengupta:2023yof}.

However, a major obstacle remains: we still lack a clear understanding of the dynamical mechanism responsible for the spontaneous breaking of Weyl conformal symmetry. While some proposals exist, a fully satisfactory explanation is yet to be found.

\section{Acknowledgments}
We thank R. Garattini for  discussions and useful comments.

\bibliographystyle{ieeetr}
\bibliography{biblio}

\begin{thebibliography}{10}

\bibitem{Morris:1988cz}
M.~S. Morris and K.~S. Thorne, ``{Wormholes in space-time and their use for interstellar travel: A tool for teaching general relativity},'' {\em Am. J. Phys.}, vol.~56, pp.~395--412, 1988.

\bibitem{Visser:1995cc}
M.~Visser, {\em {Lorentzian wormholes: From Einstein to Hawking}}.
\newblock 1995.

\bibitem{Einstein:1935tc}
A.~Einstein and N.~Rosen, ``{The Particle Problem in the General Theory of Relativity},'' {\em Phys. Rev.}, vol.~48, pp.~73--77, 1935.

\bibitem{Ellis:1973yv}
H.~G. Ellis, ``{Ether flow through a drainhole - a particle model in general relativity},'' {\em J. Math. Phys.}, vol.~14, pp.~104--118, 1973.

\bibitem{Hawking:1991nk}
S.~W. Hawking, ``{The Chronology protection conjecture},'' {\em Phys. Rev. D}, vol.~46, pp.~603--611, 1992.

\bibitem{Visser:1992tx}
M.~Visser, ``{From wormhole to time machine: Comments on Hawking's chronology protection conjecture},'' {\em Phys. Rev. D}, vol.~47, pp.~554--565, 1993.

\bibitem{Bronnikov:1973fh}
K.~A. Bronnikov, ``{Scalar-tensor theory and scalar charge},'' {\em Acta Phys. Polon. B}, vol.~4, pp.~251--266, 1973.

\bibitem{Visser:1989kh}
M.~Visser, ``{Traversable wormholes: Some simple examples},'' {\em Phys. Rev. D}, vol.~39, pp.~3182--3184, 1989.

\bibitem{Hochberg:1998ha}
D.~Hochberg and M.~Visser, ``{Dynamic wormholes, anti-trapped surfaces, and energy conditions},'' {\em Phys. Rev. D}, vol.~58, p.~044021, 1998.

\bibitem{Hochberg:1998ii}
D.~Hochberg and M.~Visser, ``{The Null energy condition in dynamic wormholes},'' {\em Phys. Rev. Lett.}, vol.~81, pp.~746--749, 1998.

\bibitem{Lobo:2004rp}
F.~S.~N. Lobo, ``{Energy conditions, traversable wormholes and dust shells},'' {\em Gen. Rel. Grav.}, vol.~37, pp.~2023--2038, 2005.

\bibitem{Lobo:2005us}
F.~S.~N. Lobo, ``{Phantom energy traversable wormholes},'' {\em Phys. Rev. D}, vol.~71, p.~084011, 2005.

\bibitem{Garattini:2007ff}
R.~Garattini and F.~S.~N. Lobo, ``{Self sustained phantom wormholes in semi-classical gravity},'' {\em Class. Quant. Grav.}, vol.~24, pp.~2401--2413, 2007.

\bibitem{Bouhmadi-Lopez:2014gza}
M.~Bouhmadi-L\'opez, F.~S.~N. Lobo, and P.~Mart\'\i{}n-Moruno, ``{Wormholes minimally violating the null energy condition},'' {\em JCAP}, vol.~11, p.~007, 2014.

\bibitem{Garattini:2019ivd}
R.~Garattini, ``{Casimir Wormholes},'' {\em Eur. Phys. J. C}, vol.~79, no.~11, p.~951, 2019.

\bibitem{Lobo:2020ffi}
F.~S.~N. Lobo, M.~E. Rodrigues, M.~V. de~Sousa~Silva, A.~Simpson, and M.~Visser, ``{Novel black-bounce spacetimes: wormholes, regularity, energy conditions, and causal structure},'' {\em Phys. Rev. D}, vol.~103, no.~8, p.~084052, 2021.

\bibitem{Barros:2023pre}
B.~J. Barros, A.~de~la Cruz-Dombriz, and F.~S.~N. Lobo, ``{Wormholes with matter haunted by conformally coupled ghosts},'' {\em Phys. Rev. D}, vol.~108, no.~8, p.~084028, 2023.

\bibitem{Garattini:2023kmr}
R.~Garattini and A.~G. Tzikas, ``{Traversable wormholes induced by stress energy conservation: combining Casimir energy with a scalar~field},'' {\em JCAP}, vol.~12, p.~019, 2024.

\bibitem{Garattini:2024jkr}
R.~Garattini and M.~Faizal, ``{Hot Casimir wormholes},'' {\em JCAP}, vol.~01, no.~081, p.~081, 2025.

\bibitem{Maldacena:2013xja}
J.~Maldacena and L.~Susskind, ``{Cool horizons for entangled black holes},'' {\em Fortsch. Phys.}, vol.~61, pp.~781--811, 2013.

\bibitem{Penington:2019kki}
G.~Penington, S.~H. Shenker, D.~Stanford, and Z.~Yang, ``{Replica wormholes and the black hole interior},'' {\em JHEP}, vol.~03, p.~205, 2022.

\bibitem{Almheiri:2019qdq}
A.~Almheiri, T.~Hartman, J.~Maldacena, E.~Shaghoulian, and A.~Tajdini, ``{Replica Wormholes and the Entropy of Hawking Radiation},'' {\em JHEP}, vol.~05, p.~013, 2020.

\bibitem{Almheiri:2020cfm}
A.~Almheiri, T.~Hartman, J.~Maldacena, E.~Shaghoulian, and A.~Tajdini, ``{The entropy of Hawking radiation},'' {\em Rev. Mod. Phys.}, vol.~93, no.~3, p.~035002, 2021.

\bibitem{Damour:2007ap}
T.~Damour and S.~N. Solodukhin, ``{Wormholes as black hole foils},'' {\em Phys. Rev. D}, vol.~76, p.~024016, 2007.

\bibitem{Cardoso:2016rao}
V.~Cardoso, E.~Franzin, and P.~Pani, ``{Is the gravitational-wave ringdown a probe of the event horizon?},'' {\em Phys. Rev. Lett.}, vol.~116, no.~17, p.~171101, 2016.
\newblock [Erratum: Phys.Rev.Lett. 117, 089902 (2016)].

\bibitem{Konoplya:2016hmd}
R.~A. Konoplya and A.~Zhidenko, ``{Wormholes versus black holes: quasinormal ringing at early and late times},'' {\em JCAP}, vol.~12, p.~043, 2016.

\bibitem{Harko:2017fra}
T.~Harko, Z.~Kov\'acs, and F.~S.~N. Lobo, ``{Astrophysical Signatures of Thin Accretion Disks in Wormhole Spacetimes},'' {\em Fundam. Theor. Phys.}, vol.~189, pp.~63--88, 2017.

\bibitem{Bueno:2017hyj}
P.~Bueno, P.~A. Cano, F.~Goelen, T.~Hertog, and B.~Vercnocke, ``{Echoes of Kerr-like wormholes},'' {\em Phys. Rev. D}, vol.~97, no.~2, p.~024040, 2018.

\bibitem{Cardoso:2019rvt}
V.~Cardoso and P.~Pani, ``{Testing the nature of dark compact objects: a status report},'' {\em Living Rev. Rel.}, vol.~22, no.~1, p.~4, 2019.

\bibitem{Akil:2022coa}
A.~Akil, M.~Cadoni, L.~Modesto, M.~Oi, and A.~P. Sanna, ``{Semiclassical spacetimes at super-Planckian scales from delocalized sources},'' {\em Phys. Rev. D}, vol.~108, no.~4, p.~044051, 2023.

\bibitem{Hawking:1988ae}
S.~W. Hawking, ``{Wormholes in Space-Time},'' {\em Phys. Rev. D}, vol.~37, pp.~904--910, 1988.

\bibitem{Vollick:1998qf}
D.~N. Vollick, ``{Wormholes in string theory},'' {\em Class. Quant. Grav.}, vol.~16, pp.~1599--1604, 1999.

\bibitem{Lobo:2007qi}
F.~S.~N. Lobo, ``{A General class of braneworld wormholes},'' {\em Phys. Rev. D}, vol.~75, p.~064027, 2007.

\bibitem{Arkani-Hamed:2007cpn}
N.~Arkani-Hamed, J.~Orgera, and J.~Polchinski, ``{Euclidean wormholes in string theory},'' {\em JHEP}, vol.~12, p.~018, 2007.

\bibitem{Garattini:2008xz}
R.~Garattini and F.~S.~N. Lobo, ``{Self-sustained traversable wormholes in noncommutative geometry},'' {\em Phys. Lett. B}, vol.~671, pp.~146--152, 2009.

\bibitem{Garattini:2011fs}
R.~Garattini and F.~S.~N. Lobo, ``{Self-sustained wormholes in modified dispersion relations},'' {\em Phys. Rev. D}, vol.~85, p.~024043, 2012.

\bibitem{Sengupta:2021wvi}
R.~Sengupta, S.~Ghosh, M.~Kalam, and S.~Ray, ``{Traversable wormhole on the brane with non-exotic matter: a broader view},'' {\em Class. Quant. Grav.}, vol.~39, no.~10, p.~105004, 2022.

\bibitem{Sengupta:2023yof}
R.~Sengupta, S.~Ghosh, and M.~Kalam, ``{Lorentzian wormhole in the framework of loop quantum cosmology},'' {\em Eur. Phys. J. C}, vol.~83, no.~9, p.~830, 2023.

\bibitem{Cadoni:1994av}
M.~Cadoni and M.~Cavaglia, ``{Cosmological and wormhole solutions in low-energy effective string theory},'' {\em Phys. Rev. D}, vol.~50, pp.~6435--6443, 1994.

\bibitem{Cadoni:1995pg}
M.~Cadoni and M.~Cavaglia, ``{Instability of the R**3 x S**1 vacuum in low-energy effective string theory},'' {\em Phys. Rev. D}, vol.~52, pp.~2583--2586, 1995.

\bibitem{Gonzalez:2008wd}
J.~A. Gonzalez, F.~S. Guzman, and O.~Sarbach, ``{Instability of wormholes supported by a ghost scalar field. I. Linear stability analysis},'' {\em Class. Quant. Grav.}, vol.~26, p.~015010, 2009.

\bibitem{Bronnikov:2012ch}
K.~A. Bronnikov, R.~A. Konoplya, and A.~Zhidenko, ``{Instabilities of wormholes and regular black holes supported by a phantom scalar field},'' {\em Phys. Rev. D}, vol.~86, p.~024028, 2012.

\bibitem{Reuter:2008qx}
M.~Reuter and H.~Weyer, ``{Conformal sector of Quantum Einstein Gravity in the local potential approximation: Non-Gaussian fixed point and a phase of unbroken diffeomorphism invariance},'' {\em Phys. Rev. D}, vol.~80, p.~025001, 2009.

\bibitem{tHooft:2010xlr}
G.~'t~Hooft, ``{Probing the small distance structure of canonical quantum gravity using the conformal group},'' 9 2010.

\bibitem{tHooft:2010mvw}
G.~'t~Hooft, ``{The Conformal Constraint in Canonical Quantum Gravity},'' 11 2010.

\bibitem{Mannheim:2011ds}
P.~D. Mannheim, ``{Making the Case for Conformal Gravity},'' {\em Found. Phys.}, vol.~42, pp.~388--420, 2012.

\bibitem{Kallosh:2013lkr}
R.~Kallosh and A.~Linde, ``{Superconformal generalizations of the Starobinsky model},'' {\em JCAP}, vol.~06, p.~028, 2013.

\bibitem{tHooft:2015vaz}
G.~'t~Hooft, ``{Local conformal symmetry: The missing symmetry component for space and time},'' {\em Int. J. Mod. Phys. D}, vol.~24, no.~12, p.~1543001, 2015.

\bibitem{Rachwal:2018gwu}
L.~Rachwa\l{}, ``{Conformal Symmetry in Field Theory and in Quantum Gravity},'' {\em Universe}, vol.~4, no.~11, p.~125, 2018.

\bibitem{Modesto:2016max}
L.~Modesto and L.~Rachwal, ``{Finite Conformal Quantum Gravity and Nonsingular Spacetimes},'' 5 2016.

\bibitem{Rachwal:2022pfe}
L.~Rachwa\l{}, ``{Introduction to Quantization of Conformal Gravity},'' {\em Universe}, vol.~8, no.~4, p.~225, 2022.

\bibitem{Giacometti:2024qva}
G.~Giacometti, A.~Bonanno, S.~Plumari, and D.~Zappal\`a, ``{Spontaneous breaking of diffeomorphism invariance in conformally reduced quantum gravity},'' 10 2024.

\bibitem{Hohmann:2018shl}
M.~Hohmann, C.~Pfeifer, M.~Raidal, and H.~Veerm\"ae, ``{Wormholes in conformal gravity},'' {\em JCAP}, vol.~10, p.~003, 2018.

\bibitem{Teo:1998dp}
E.~Teo, ``{Rotating traversable wormholes},'' {\em Phys. Rev. D}, vol.~58, p.~024014, 1998.

\bibitem{Kuhfittig:2003wr}
P.~K.~F. Kuhfittig, ``{Axially symmetric rotating traversable wormholes},'' {\em Phys. Rev. D}, vol.~67, p.~064015, 2003.

\bibitem{Kashargin:2007mm}
P.~E. Kashargin and S.~V. Sushkov, ``{Slowly rotating wormholes: The First order approximation},'' {\em Grav. Cosmol.}, vol.~14, pp.~80--85, 2008.

\bibitem{Kashargin:2008pk}
P.~E. Kashargin and S.~V. Sushkov, ``{Slowly rotating scalar field wormholes: The Second order approximation},'' {\em Phys. Rev. D}, vol.~78, p.~064071, 2008.

\bibitem{Dzhunushaliev:2013jja}
V.~Dzhunushaliev, V.~Folomeev, B.~Kleihaus, J.~Kunz, and E.~Radu, ``{Rotating Wormholes in Five Dimensions},'' {\em Phys. Rev. D}, vol.~88, p.~124028, 2013.

\bibitem{Kleihaus:2014dla}
B.~Kleihaus and J.~Kunz, ``{Rotating Ellis Wormholes in Four Dimensions},'' {\em Phys. Rev. D}, vol.~90, p.~121503, 2014.

\bibitem{Franzin:2022iai}
E.~Franzin, S.~Liberati, J.~Mazza, R.~Dey, and S.~Chakraborty, ``{Scalar perturbations around rotating regular black holes and wormholes: Quasinormal modes, ergoregion instability, and superradiance},'' {\em Phys. Rev. D}, vol.~105, no.~12, p.~124051, 2022.

\bibitem{Garattini:2025gfq}
R.~Garattini and A.~G. Tzikas, ``{Rotating Casimir Wormholes},'' {\em arXiv:2502.19995}, 2025.

\bibitem{Modesto:2021yyf}
L.~Modesto, T.~Zhou, and Q.~Li, ``{Geometric Origin of the Galaxies\textquoteright{} Dark Side},'' {\em Universe}, vol.~10, no.~1, p.~19, 2024.

\bibitem{Bambi:2016wdn}
C.~Bambi, L.~Modesto, and L.~Rachwa\l, ``{Spacetime completeness of non-singular black holes in conformal gravity},'' {\em JCAP}, vol.~05, p.~003, 2017.

\bibitem{Chakrabarty:2017ysw}
H.~Chakrabarty, C.~A. Benavides-Gallego, C.~Bambi, and L.~Modesto, ``{Unattainable extended spacetime regions in conformal gravity},'' {\em JHEP}, vol.~03, p.~013, 2018.

\bibitem{Bach:1921zdq}
R.~Bach, ``{Zur Weylschen Relativit\"atstheorie und der Weylschen Erweiterung des Kr\"ummungstensorbegriffs},'' {\em Math. Z.}, vol.~9, no.~1, pp.~110--135, 1921.

\bibitem{Mannheim:1988dj}
P.~D. Mannheim and D.~Kazanas, ``{Exact Vacuum Solution to Conformal Weyl Gravity and Galactic Rotation Curves},'' {\em Astrophys. J.}, vol.~342, pp.~635--638, 1989.

\bibitem{Wheeler:2013ora}
J.~T. Wheeler, ``{Weyl gravity as general relativity},'' {\em Phys. Rev. D}, vol.~90, no.~2, p.~025027, 2014.

\bibitem{CALLAN197042}
C.~G. Callan, S.~Coleman, and R.~Jackiw, ``A new improved energy-momentum tensor,'' {\em Annals of Physics}, vol.~59, no.~1, pp.~42--73, 1970.

\bibitem{Percacci:2011uf}
R.~Percacci, ``{Renormalization group flow of Weyl invariant dilaton gravity},'' {\em New J. Phys.}, vol.~13, p.~125013, 2011.

\bibitem{Marolf:2021kjc}
D.~Marolf and J.~E. Santos, ``{AdS Euclidean wormholes},'' {\em Class. Quant. Grav.}, vol.~38, no.~22, p.~224002, 2021.

\end{thebibliography}

\newpage 

\end{document}